\documentclass[onecolumn,secnumarabic,amssymb, nobibnotes, aps, pr,superscriptaddress]{revtex4-1}
\usepackage{amsmath}

\usepackage{times}
\usepackage{bm}
\usepackage{natbib}

\usepackage[plain,noend]{algorithm2e}

\usepackage{stackengine,graphicx}

\def\l{\left}
\def\r{\right}

\newcommand{\f}{\frac}

\begin{document}

\title{Crossing points in survival analysis sensitively depend on system conditions}

\author{Thomas~McAndrew, Ph.D}
\email{tmcandrew@crf.org}
\affiliation{Cardiovascular Research Foundation}

\author{Bjorn~Redfors, M.D. Ph.D}
\affiliation{Cardiovascular Research Foundation}
\affiliation{Department of Cardiology Sahlgrenska University Hospital, Gothenburg, Sweden}

\author{Yiran~Zhang, M.S.}
\affiliation{Cardiovascular Research Foundation}

\author{Aaron~Crowley, M.A.}
\affiliation{Cardiovascular Research Foundation}

\author{Shmuel~Chen, M.D., Ph.D.}
\affiliation{Cardiovascular Research Foundation}
\affiliation{Hadassah Medical Center Jerusalem, Israel}

\author{Gregg~Stone, M.D.}
\affiliation{Cardiovascular Research Foundation}
\affiliation{Columbia University Medical Center}

\author{Paul~Jenkins, Ph.D.}
\affiliation{Bassett Research Institute}

\date{2018-05-07}

\begin{abstract}
  Crossing survival curves complicate how we interpret results from a clinical trial's primary endpoint.
%
We find the function to determine a crossing point's location depends exponentially on individual survival curves.
  This exponential relationship between survival curves and the crossing point transforms small survival curve errors into large crossing point errors.
  In most cases, crossing points are sensitive to individual survival errors and may make accurately locating a crossing point unsuccessful.
  We argue more complicated analyses for mitigating crossing points should be reserved only after first exploring a crossing point's variability, or hypothesis tests account for crossing point variability.
\end{abstract}

\maketitle

\section{Introduction}
Crossing survival curves challenge how a treatment benefits patients compared to control therapy~\citep{stone2016everolimus,king2014phase,velazquez2016coronary,montgomery2011desensitization,united2010endovascular}.
While past research explores making decisions in light of crossing survival curves~\citep{qiu2008two,chen2017improved,fleming1980modified,mantel1988crossing,logan2008comparing,li2015statistical,bouliotis2011crossing,yang2010improved,zucker1990weighted,yang2005semiparametric}, little research explores crossing point uncertainty.
We derive a crossing point equation and discover how small errors estimating two survival curves propagate to large differences in crossing points. 


Past work adapts routine statistical tests to manage crossing survival curves.
Supremum methods build statistics around the largest survival difference between arms~\citep{fleming1980modified,mantel1988crossing}.
Combination techniques estimate and join survival differences before and after an assumed crossing point~\citep{qiu2008two,chen2017improved}.
Weighting methods apply weights to events that occur before or after a crossing point, and typically shift attention away from survival differences before a crossing point and towards later survival differences~\citep{yang2010improved,zucker1990weighted}.
Newer statistical techniques, such as restricted mean survival~\citep{royston2011use}, study different survival attributes that don't as easily fall prey to crossing points.
Past work spends time mitigating the effect of crossing points without specifically addressing how two survival curves cross.

In the following, we study (i) where two Weibull-distributed curves cross and define an analytical expression for a crossing point, (ii) how errors estimating each survival parameter affect errors estimating a crossing point, (iv) determine survival curve characteristics that lead to difficulty estimating a crossing point, and (v) compare parameter estimate errors versus crossing point errors as a function of sample size.  

\section{Methods}

We derive a formula identifying when two survival curves cross $(t_{\chi})$ and establish this function sensitively depends on all four survival curve parameters (2 parameters for the treatment population and 2 parameters for the control population).
We find errors demonstrate power law and exponential properties~(Fig.\ref{fig1.txError}), increasing similarity between survival curves increases crossing point uncertainty~(Fig.\ref{fig2.fromL2tx} \&~Fig.\ref{fig3.fromK2tx}), and while an increasing sample size decreases survival curve parameter errors at a similar rate, crossing point errors depend on survival curve characteristics~(Fig.\ref{fig4.errFuncOfN}). 

\subsection{Crossing-point equation}

Consider each treatment arm's survival distributed Weibull with failure distributed
\begin{align}
  p(T<t|\lambda,k) = \alpha(t|\lambda,k) = e^{-(\lambda t)^{k}}. \nonumber
\end{align}

We find where two survival curves (one from the treatment group $\alpha_{1}(t)$ and a second from the control group $\alpha_{0}(t)$) cross by setting them equal and solving for $t$.
The crossing point ($t_{\chi}$) equals
\begin{equation}
  t_{\chi} = \exp \l(- \f{k_{1} \log(\lambda_{1}) - k_{0}\log(\lambda_{0})}{k_{1} - k_{0}} \r) \label{crossPoint} 
\end{equation}
where $\lambda_{x}$ represents the survival's failure parameter, $k_{x}$ represents a survival's shape parameter, and we identify the treatment group as $x=1$ and control group as $x=0$.
This equation shows us two survival curves (i) will not cross with equal shapes $k$, (ii) will not cross with divergent failures $\lambda$, and (iii) errors in either survival curve will propagate to errors in the crossing point's location.

\subsection{Crossing point sensitivity}

We derived a simple formula relating a crossing point to individual survival curves, and study how errors in estimating a survival curve can cause errors locating a crossing point. The $\epsilon$ error a parameter $p$ contributes to a relative error in $t_{\chi}$ is defined as
\begin{align}
  \mathcal{R}(\epsilon | p) = t_{\chi}(\epsilon)/t_{\chi} - 1. \label{relDiff}
\end{align}
We can determine how sensitively a crossing point depends on $\lambda$ and $k$ and fundamental quantities related to crossing point sensitivity by studying how crossing point errors scale with these two parameters.

Using \eqref{relDiff} and relative to $\lambda$, we find $t_{\chi}$'s error scales like
\begin{align}
  \mathcal{R}(\phi | \lambda) \sim \l \lvert \l(\f{1}{1+\phi}\r)^{\l(1- \f{k_{0}}{k_{1}}\r)^{-1}} \r \rvert \nonumber 
\end{align}
where $\phi$ represents a percent error in $\lambda$ (defined as $\epsilon = \phi \times \lambda$).
We see $\lambda$'s influence on $t_{\chi}$ scales like a power law with exponent equal to the relative difference between survival curve shapes $1- \f{k_{0}}{k_{1}}$.
A larger relative difference between $k$s shrinks $\lambda$'s influence on $t_{\chi}$.

Relative to $k$, a crossing point's error scales like
\begin{align}
  \mathcal{R}(\phi | k) \sim \l \lvert \l(\f{\lambda_{1}}{\lambda_{0}}\r)^{ \phi \l[\l(1-\f{k_0}{k_1}\r)\l(\f{k_1}{k_0}- 1\r)\r]^{-1}} \r \rvert \nonumber 
\end{align}
where $\phi$ represents a percent error in $k$.
Errors from $k$ depend on the relative difference between shape parameters and the ratio between failures $\lambda_{1}$ and $\lambda_{0}$.
We found two quantities, the relative difference between shape parameters $\l(1 - \f{k_{0}}{k_{1}}\r)$ and ratio of failures $\l(\f{\lambda_{1}}{\lambda_{0}}\r)$, play key roles in generating crossing point errors.

\begin{figure}[ht!]
  \centering
  \includegraphics[scale=0.5]{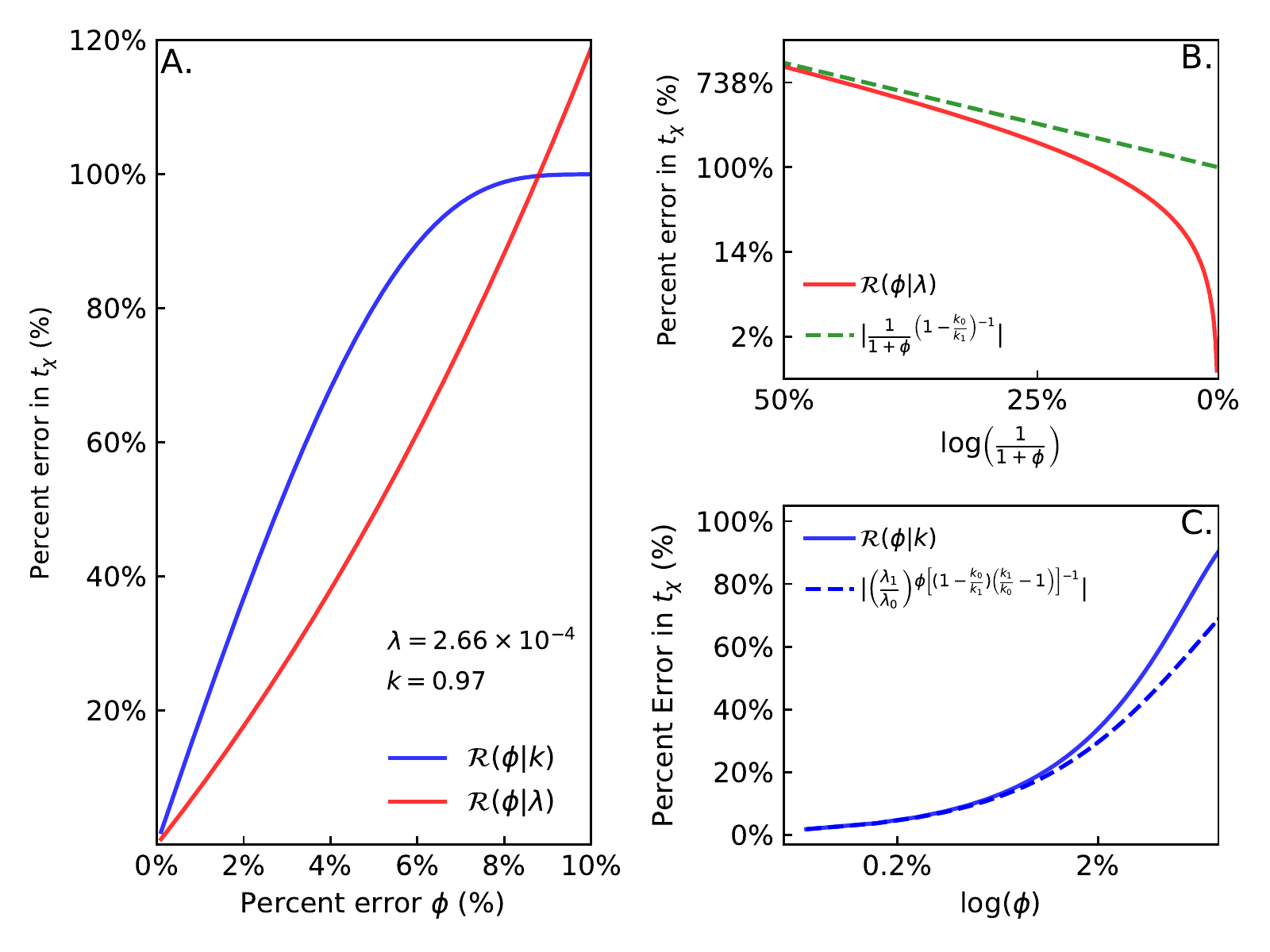}
  \caption{We see the crossing point $(t_{\chi})$ function~\eqref{crossPoint} amplifies the shape parameter ($k$) and failure parameter ($\lambda$)'s estimation errors~(A.), and the relative difference between survival curve shapes~(B.) and ratio between failures~(C.) key components driving crossing point error. \label{fig1.txError}}
\end{figure}

Considering a simple empirical example with the treatment group's $365$ $(730)$ day failure rate equal to 10\% (18\%) and control group's $365$ $(730)$ day failure rate equal to 10\% (20\%), we see small errors made estimating both $\lambda$ and $k$ cause large errors locating a crossing point~(Fig.\ref{fig1.txError}A.).
Producing $\lambda$ or $k$ values 10\% different than the true parameter values could inflate $t_{\chi}$'s error by as much as 100\%.
Exploring how errors in $\lambda$ relate to $t_{\chi}$~(Fig.\ref{fig1.txError}B.), this power law relationship quickly inflates small $\lambda$ errors into notable $t_{\chi}$ errors.
Measurement errors in $k$ exponentially amplify $t_{\chi}$ errors~(Fig.\ref{fig1.txError}C.).

Errors from $\lambda$ and $k$ to $t_{\chi}$ depend on properties of our system.
We can consider the errors from $\lambda$ to $t_{\chi}$ and from $k$ to $t_{\chi}$ as functions of these system level properties.

\begin{figure}[ht!]
  \centering
  \includegraphics[scale=0.50]{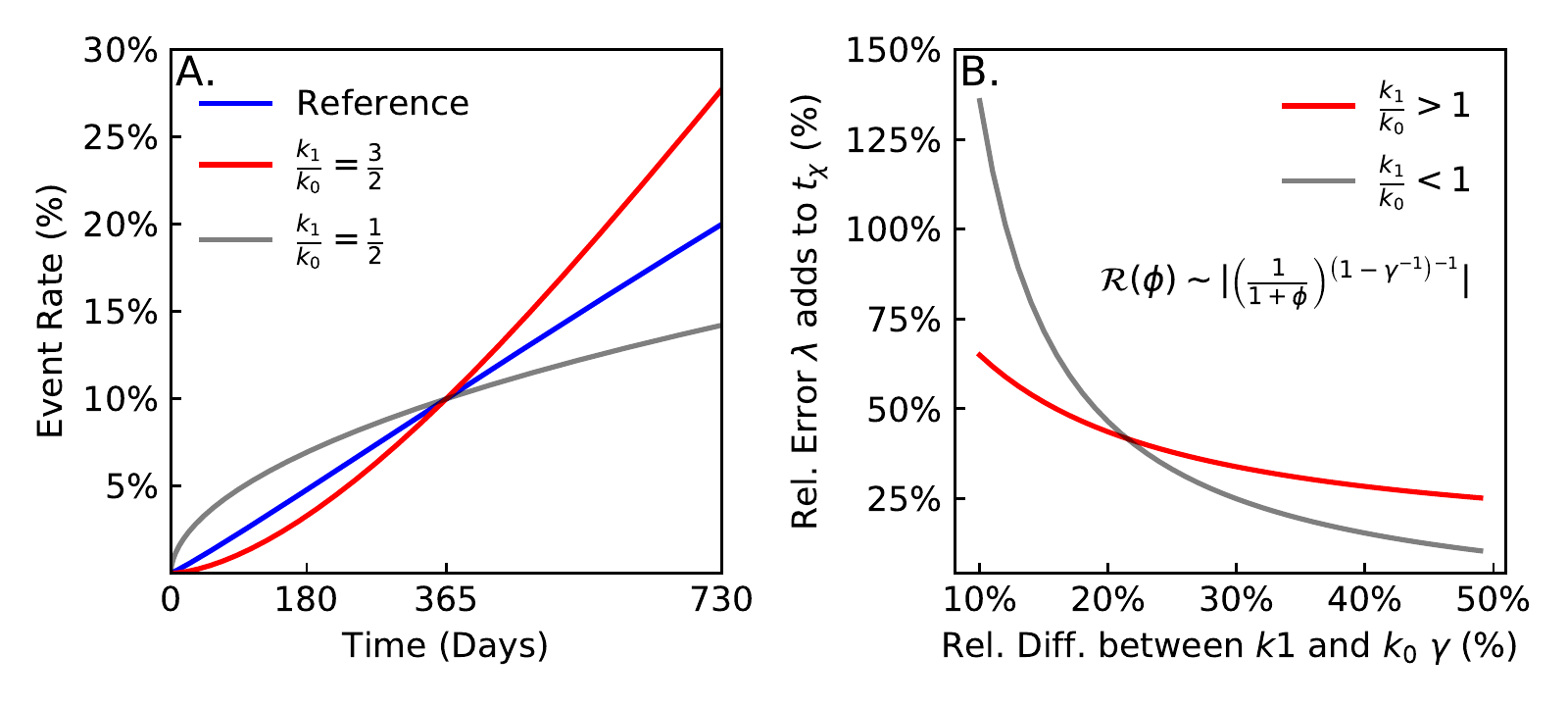}
  \caption{Increasing the relative difference between shape parameters ($k$s) causes the two survival curves to cross at a steeper angle~(A.).
           The relative error the failure parameter ($\lambda$) contributes to the crossing point $(t_{\chi})$ decreases as the relative difference between shapes increases~(B.).
           \label{fig2.fromL2tx}}
       \end{figure}

We consider $\lambda$'s error function as 
\begin{align}
  \mathcal{R}(\gamma|\phi,\lambda) \sim \l \lvert \l(\f{1}{1+\phi}\r)^{\l(1 - \f{1}{\gamma}\r)^{-1}} \r \rvert, \nonumber
\end{align}
depending on $\gamma = \f{k_{1}}{k_{0}}$ and fixed $\phi$.
Increasing $\gamma$ away from $1$ decreases $t_{\chi}$'s error due to $\phi$ (errors in $\lambda$)~(Fig.\ref{fig2.fromL2tx}B.).
A larger relative difference between shape parameters causes the two survival curves to meet at a steeper angle~(Fig.\ref{fig2.fromL2tx}A.) and dampen errors.  

\begin{figure}[ht!]
  \centering
  \includegraphics[scale=0.5]{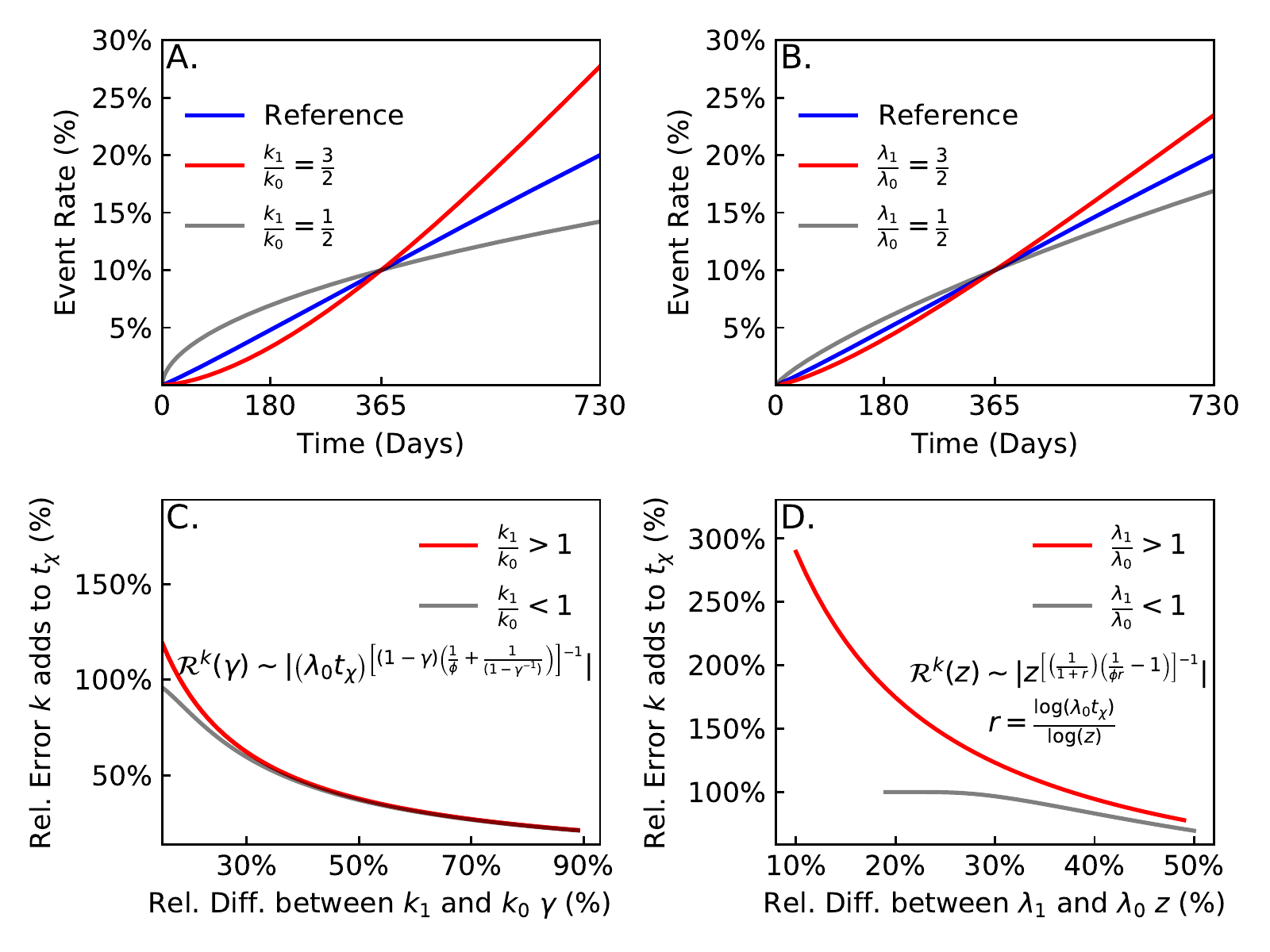}
  \caption{ Contrasting $\lambda$s and $k$s between survival curves lower the error $k$ transfers to $t_{\chi}$. 
            A larger difference between $k$s result in more sudden crossings~(A.).
            Larger difference between survival shapes decreases the relative error $k$ contributes to $t_{\chi}$~(C.).
            Larger differences between $\lambda$s also force survival curves to cross at sharper angles~(B.) and decrease the relative error $k$ contributes to $t_{\chi}$~(D.)
    \label{fig3.fromK2tx}}
\end{figure}

We consider $k$'s error function as 
\begin{align}
  \mathcal{R}(\gamma|\phi,k) \sim \l \lvert \l(\lambda_{0}t_{\chi}\r)^{\l[ \l(1-\gamma \r) \l( \f{1}{\phi} + \f{1}{1+\gamma^{-1}} \r) \r]^{-1}} \r \rvert \nonumber
\end{align}
depending on $\gamma$, and decreasing steeply as $\gamma$ moves away from $1$~(Fig.\ref{fig3.fromK2tx}C.) or as the two survival curve shapes become more distinct~(Fig.\ref{fig3.fromK2tx}A.).
We can also consider $k$'s error function as 
\begin{align}
  \mathcal{R}(z|\phi,k) &\sim \l \lvert z^{\l[ \l(\f{1}{1+r}\r) \l(\f{1}{\phi r} - 1\r)\r]^{-1}} \r \rvert\\ \nonumber
  r &= \f{\log\l(\lambda_{0}t_{\chi}\r)}{\log(z)}, \nonumber
\end{align}
with $z = \f{\lambda_{1}}{\lambda_{0}}$, and decreasing as $z$ moves away from $1$~(Fig.\ref{fig3.fromK2tx}D.) or as the patient's event probability separates~(Fig.\ref{fig3.fromK2tx}B.).

We find the underlying properties of the patient population and event under study controls how well we can estimate a crossing point.
More contrasted survival curve shapes or survival curve failure rates allow us to better estimate crossing points.
Even with dissimilar survival curve properties, small changes in survival curve parameter cause large swings in where two curves cross.
This crossing point sensitivity will manifest from drawing a finite sample and estimating individual parameters for treatment and control patient populations.

\section{Statistical Inference}
With a data sample, we first estimate each survival curve's shape and scale, and second estimate the crossing point.
Assuming the Weibull distribution generated our survival data, we estimate the shape and failure's posterior probability as
\begin{equation}
  p(\lambda, k | \mathcal{D}) \propto p(\mathcal{D} | \lambda, k ) \times p(\lambda, k), \nonumber
\end{equation}
or log posterior probability
\begin{multline}
  \log \l( p(\lambda, k) \r) \propto \l(E + \alpha - 1 \r) \log(k) + (k E + \gamma - 1)\log(\lambda) + (k-1)\sum_{i=1}^{E}\log(t_{i}) - \l( \lambda^{k} \sum_{j=1}^{N} t_{i}^{k} + \beta k + \eta \lambda \r) \nonumber
\end{multline}
with $E$ events, $N$ patients, $N-E$ censored observations, gamma-distributed prior probability $p(k) = \dfrac{\beta^{\alpha}}{\Gamma(\alpha)}k^{\alpha-1}e^{-\beta k}$, and gamma-distributed prior probability $p(\lambda) = \dfrac{\eta^{\gamma}}{\Gamma(\gamma)} k^{\gamma-1}e^{- \eta k}$.

We simulated $N$ control patients and $N$ treatment patients from $N=200$ to $N = 1900$.
For each sample size, our control population's shape equaled $k_{0}=1.08$ and failure equaled $\lambda_{0} = 3.43 \times 10^{-4}$.
The treatment population's shape and failure was set to four scenarios; two scenarios set the relative difference between treatment and control failures to $25$\% and $50$\%, and the remaining two scenarios set the relative difference between treatment and control's shape to $25$\% and $50$\%.
We fixed the crossing point for all the above scenarios to $t_{\chi} = 365$ days.

\begin{figure}[ht!]
  \centering
  \includegraphics[scale=0.5]{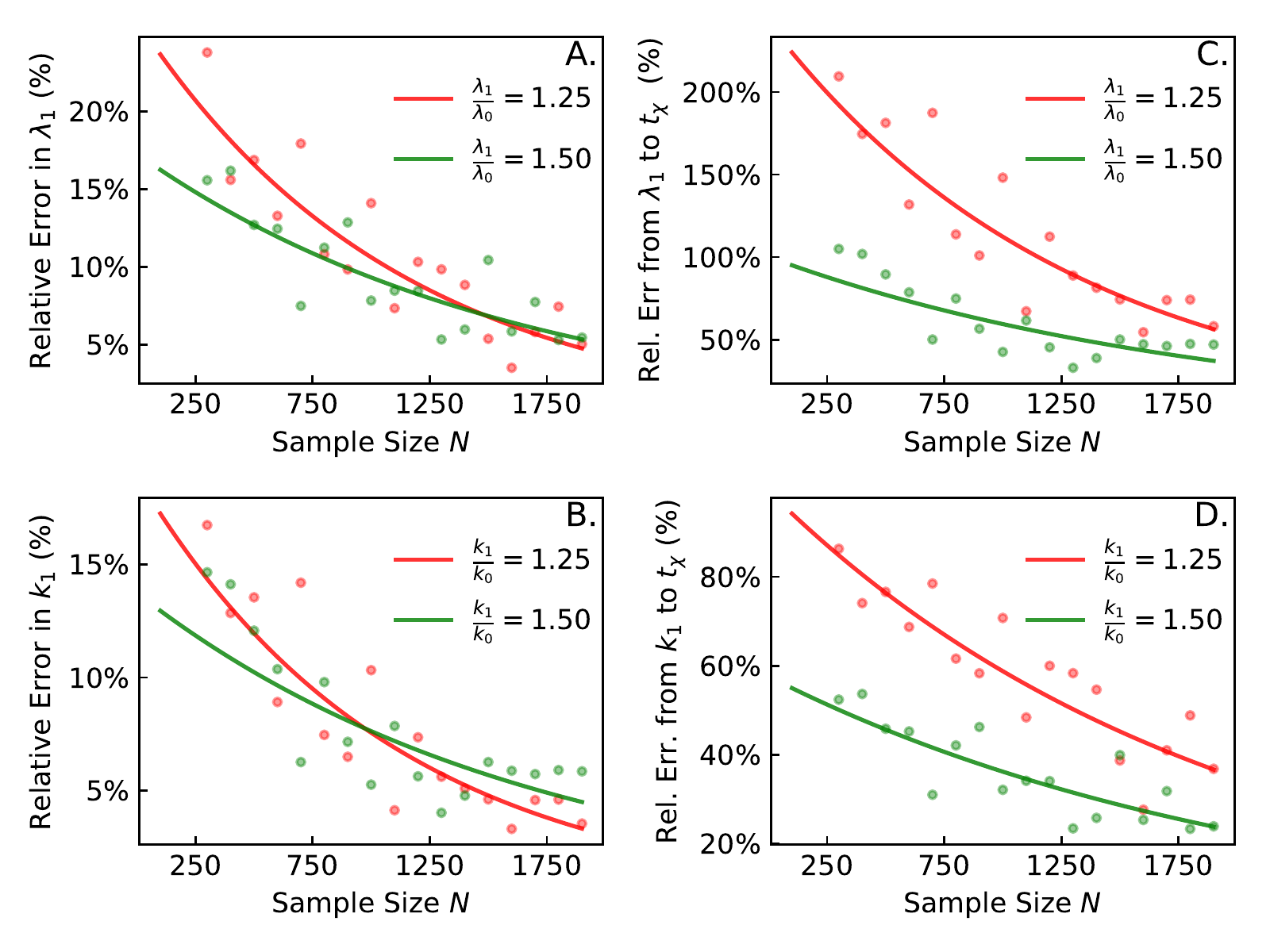}
  \caption{
    As our sample size grows, uncertainty in parameters decreases independently of their true values~(A. \& C.), while the error locating the crossing point depends on the true shape and failure values~(B. \& D.). 
    (A.) A 10-simulation average relative error in $\lambda$ from 200 to 1900 patients.
    (B.) The crossing point's average relative error caused by $\lambda$. 
    (C.) A 10-simulation average relative error in $k$ from 200 to 1900 patients.
    (D.) The crossing point's average relative error caused by $k$. 
    \label{fig4.errFuncOfN}}
\end{figure}

Although our parameter estimates relative error shrinks at the same rate regardless of the true $(\lambda_{1}, k_{1})$, the crossing point's relative error depends on the true $(\lambda_{1}, k_{1})$ and increases when either the true $k_{1}$ lies closer to $k_{0}$ or true $\lambda_{1}$ lies closer to $\lambda_{0}$. 
As our sample size $N$ grows from $200$ patients to $1900$ patents, we find both $\lambda$ and $k$'s relative error quickly shrinks toward zero percent independent of $\lambda$ or $k$'s true value~(Fig.\ref{fig4.errFuncOfN}A. and Fig.\ref{fig4.errFuncOfN}B.).
Opposing this parameter-independent shrinking, the crossing point's estimated error depends on the $\lambda$ and $k$'s true value.
Supporting our above theory, we find lower relative errors in the crossing point with more separate failures (Fig.\ref{fig4.errFuncOfN}B.) and shapes~(Fig.\ref{fig4.errFuncOfN}D.).

\section{Discussion}

The event and population we study influence $(\lambda,k)$ and determine how certain we can estimate a crossing point.
Two survival curves with similar shapes or failure parameters amplify small errors into considerable crossing point errors.
Before using more sophisticated hypothesis tests, one should consider how the event type and patients under study affect crossing point sensitivity and impact conclusions from hypothesis tests.

Previous work does not consider crossing point uncertainties and this may cause overconfident results from hypothesis tests.
If we consider the probability each event in a sample occurs after a crossing point, previous work assigns either a probability of $0$ or $1$. 
Weighting methods too conservatively devalue events before the assumed crossing point (assigning them probability $0$) and over-value events after the assumed crossing point (assigning them probability $1$).  
Combination methods that analyze survival differences before and after an assumed crossing point make similar oversights.
Choosing a specific point to analyze survival differences could underestimate variances by not accounting for events moving between groups caused by an uncertain (moving) crossing point.

This crossing point analysis was limited to Weibull-like survival curves.
Weibull distributions can flexibly model events with increasing or decreasing hazards through time, but cannot handle more complicated hazard functions.
Composite endpoints pose a significant challenge due to a mixture of early and late events, and the above may not appropriately handle these types of events.

Future work will focus on non-parametric alternatives to estimating crossing points (Weibull-free) and measuring the impact crossing points outside the time interval of interest have on analysis, and examining how hypothesis tests that handle crossing points over confidently make conclusions.
A non-parametric method to estimate crossing points may better handle non-Weibull distributions and composite endpoints.
While we studied crossing points within a highlighted time interval, locating where two curves cross will effect models that assume proportional hazards.
Evaluating hypothesis tests that assume an exact crossing point will allow us to develop better tests that do not draw overconfident conclusions. 

We should study survival curve characteristics and crossing point uncertainty before turning to more complicated hypothesis tests.
A more intense study of a crossing point may save investigators from complicated interpretations and lead to simpler more impactful messages.


\end{document}